# A two-stage approach for estimating the parameters of an age-group epidemic model from incidence data

Rami Yaari[1,2,†], Itai Dattner[1†*] and Amit Huppert[2,3]

1) Department of Statistics, University of Haifa, Aba Khoushy Ave. Mount Carmel, Haifa 3498838, Israel

2) Bio-statistical Unit, The Gertner Institute for Epidemiology and Health Policy Research, Chaim Sheba Medical Center, Tel Hashomer, 52621 Israel

3) School of Public Health, the Sackler Faculty of Medicine, Tel-Aviv University, 69978 Tel Aviv, Israel

† These authors contributed equally to this work
* To whom correspondence should be addressed; Email: idattner@stat.haifa.ac.il

## Abstract

Age-dependent dynamics is an important characteristic of many infectious diseases. Age-group epidemic models describe the infection dynamics in different age-groups by allowing to set distinct parameter values for each. However, such models are highly nonlinear and may have a large number of unknown parameters. Thus, parameter estimation of age-group models, while becoming a fundamental issue for both the scientific study and policy making in infectious diseases, is not a trivial task in practice. In this paper, we examine the estimation of the so called next-generation matrix using incidence data of a single entire outbreak, and extend the approach to deal with recurring outbreaks. Unlike previous studies, we do not assume any constraints regarding the structure of the matrix. A novel two-stage approach is developed, which allows for efficient parameter estimation from both statistical and computational perspectives. Simulation studies corroborate the ability to estimate accurately the parameters of the model for several realistic scenarios. The model and estimation method are applied to real data of influenza-like-illness in Israel. The parameter estimates of the key relevant epidemiological parameters and the recovered structure of the estimated next-generation matrix are in line with results obtained in previous studies.

I. Introduction

In recent years, mathematical models have proven to be an effective tool for examining and exploring the dynamics of the spread of infectious diseases (1–5). Special practical interest in such models stems from the fact that they are currently the only systematic way to study possible control and mitigation strategies (6–10). At present, a major limitation in applying such models to deal with real life policy issues is the need to assure that the values used for the various parameters in the model are realistic. Whereas some parameters can be estimated using direct measurements or previous knowledge, other parameter values can only be estimated by model fitting to data collected during an outbreak. To do so for a large number of parameters, it is necessary to have sufficient amounts of data and to develop statistical methods which will allow efficient parameter estimation, both computationally and statistically. Thus, parameter estimation by fitting mathematical models to outbreak data becomes a fundamental issue for both the scientific study and policy making in infectious diseases (11–14).

For many infectious diseases there is a clear age-dependent dynamic, as can be seen from data regarding the burden of disease in different age-groups (15–17). Age-group epidemic models describe the infection dynamics in each age-group by allowing to set distinct parameter values for each group. One common methodological approach to quantify such heterogeneities is the use of a 'who acquires infection from whom' (WAIFW) matrix, which specifies the different transmission rates within and between the different age-groups (1,18,19). A closely related concept is the next-generation matrix (NGM), which is an extension of the basic reproduction number $R_0$ for an age-group model (20,21) (see section II for details).

The inference of the WAIFW matrix or the NGM has been limited historically. In most previous studies, age-group parameters were inferred using serological data, which can be used as a proxy of the attack rate in each age-group, while assuming that the disease is at a steady state (see (1)). In this case, there is a single data point for the disease prevalence for each age-group, limiting the number of the matrix parameters which can be estimated. Efforts in the literature to bypass and simplify the problem attempt to estimate a smaller number of the matrix parameters. The first attempt in this direction was Schenzle's pioneering studies, where he introduced assumptions (constrains) regarding the structure of the WAIFW matrix (19). One possible such structure simplification was to assume proportionate mixing, or supposing that certain groups of matrix elements are equal (1,22,23). Klepac and co-workers estimated such a constrained WAIFW matrix with three age-groups using incidence data from a phocine distemper virus outbreak in seals (24). Glass and co-workers, estimated several types of constrained NGM matrices with two age-groups of children and adults using incidence data of the initial stages of the A/H1N1pdm influenza outbreak in Japan (25).

Another approach for estimating the WAIFW/NGM matrix is by combining contact tracing data together with incidence data. Katriel and co-workers developed a method and estimated the NGM for three age-groups using data collected during the initial stages of the A/H1N1pdm influenza outbreak in Israel (26). Using a simulation study, they demonstrated that the use of both datasets together enables an accurate estimation of the NGM without the need to make any assumption regarding its structure. However, there are very few studies with sufficient contact tracing data that can be used to obtain reliable estimates. Moreover, most contact tracing datasets have estimated a relatively small contact network which might not be representative of the true network and can lead to biases in the estimations.

A third approach is to assume that the WAIFW/NGM matrix is determined exclusively by the contact patterns among individuals from different age-groups, and use empirical data on social contacts in a given population to estimate the matrix (27,28). In 2008, a large survey study, known as the POLYMOD study, was conducted in eight European countries, in order to quantify different social contact patterns, including age-group contact rates (29). Since then, there is a growing number of studies that use inferred social contact data to model the age-group transmission dynamics of different diseases (30–36). Nevertheless, the social contacts approach suffers from several drawbacks. First, the contact matrices obtained from surveys are subjected to different reporting biases. Second, the surveys were conducted in only a few countries and it is not clear how serviceable the matrices are for other countries, as contact patterns are culturally dependent. Third, the contact patterns relevant for the transmission of an infectious disease are dependent on its mode of transmission (e.g., airborne vs fecal-oral), so that different contact data should be used for different diseases, and it is not straightforward what type of contact information best reflects the transmission dynamics of a certain disease. Finally, additional factors other than contact patterns can affect the age-group transmission rates, such as age-dependent susceptibility or behavioral patterns (e.g., hygiene).

In this paper we examine for the first time the possibility of estimating the NGM using incidence data of an entire outbreak, without assuming any constraints regarding the structure of the matrix. We also examine how incidence data from recurring outbreaks can be used in order to improve our estimates. This is done by developing a two-stage estimation method, which includes as an initial step a modified version of the direct estimation method (37), and as a second step, an application of maximum likelihood estimation. We demonstrate that the first step helps in reducing the computational burden and in avoiding complex nonlinear optimization issues. Further, the use of a direct estimation approach sheds new light on theoretical properties of the model such as identifiability, which leads to valuable insights. Using a simulation study, we explore under what conditions the method is able to estimate the NGM parameters. Finally, we apply our methodology to fit a two age-group model to Influenza Like Illness (ILI) data from eleven seasonal influenza outbreaks in Israel, providing estimates for the NGM and other key model parameters.

The paper is organized as follows. Section II describes the mathematical and statistical models considered in this work. In Section III the new methodological approach is explained in detail, while Section IV presents the estimation results both from simulations and real data; Section V summarizes this research with a discussion.

## II. Mathematical and statistical models

### A. The transmission model

We use the following age-group transmission model, formulated as a discrete-time, age-of-infection model:

$$(1a) \qquad i_j(t) = \frac{S_j(t-1)}{N_j} \cdot \sum_{k=1}^{m} \left( \beta_{jk} \sum_{\tau=1}^{d} P_\tau i_k(t-\tau) \right)$$

$$(1b) \qquad S_j(t) = S_j(t-1) - i_j(t)$$

This is a slightly modified model to what was used in (26), in this case the depletion of the susceptible population is not neglected, as we intend to model a full epidemic and not just its initial stages. Here, $i_j(t)$ is the number of **newly** infected from age-group $j$ ($1 \leq j \leq m$) and $S_j(t)$ is the number of susceptible individuals from age-group $j$ on day $t$. Other works use notation $t_i$ for the discrete time points. However, for ease of notation we use $t$ in the sequel, where $t = 1, \dots, T$. The model assumes infected individual remain infected for d days, so that $i_j(t)$ depends on the number of newly infected from up to d days ago. The infectiousness of an infected individual changes with the age of his infection according to the serial interval distribution $P_{1..d}$ ($\sum_{\tau=1}^{d} P_\tau = 1$) (26). $\beta$ is the next-generation matrix (NGM). In a totally susceptible population, a single infected individual from age-group $j$ infects on average $\beta_{jk}$ individuals from age-group $k$ over the period of his/her infection. However, since not all the population is susceptible, $i_j(t)$ depends on the portion of susceptible population in age-group $j$ at day $t-1$, $S_j(t-1)/N_j$, where $N_j$ is the size of age-group $j$. For more information regarding the derivation of the model see (26,38).

The basic reproductive number for the model, or the mean number of infections caused by a single infected individual in the population as a whole, is given by $R_0 = \rho(\beta)$ where $\rho(M)$ signifies the spectral radius or maximum Eigen value of a matrix $M$ (20,21). The basic reproductive number for each age-group $j$, or the mean number of infections caused by a single infected individual from age-group $j$, is defined as $R_{0j} = \sum_{l=1}^{m} \beta_{lj}$. The effective reproductive number for the model $R_e$, which is the reproductive number in a partially susceptible population, is given by:

$$(2) \qquad R_e = \rho \begin{pmatrix} \beta_{11} S_{0_1} & \cdots & \beta_{1m} S_{0_1} \\ \vdots & \ddots & \vdots \\ \beta_{m1} S_{0_m} & \cdots & \beta_{mm} S_{0_m} \end{pmatrix}$$

where $S_{0_j} \equiv S_j(0)/N_j$ is the initial fraction of susceptible individuals in age-group $j$.

We point out a few basic facts that might shed some light on the identifiability of the model considered in equation (1). Given $t = 1, \ldots, T$ discrete points of the incidence defined in equation (1a), it is possible to write the transmission model in matrix notation as follows:

$$(3) \qquad \begin{pmatrix} i_1(1) \\ \vdots \\ i_1(T) \\ \vdots \\ i_m(1) \\ \vdots \\ i_m(T) \end{pmatrix} = \begin{pmatrix} X_{11}(1) & \cdots & X_{1m}(1) & 0 & \cdots & 0 \\ \vdots & \ddots & \vdots & \vdots & \ddots & \vdots \\ X_{11}(T) & \cdots & X_{1m}(T) & 0 & \cdots & 0 \\ & & & \ddots & & \\ 0 & \cdots & 0 & X_{m1}(1) & \cdots & X_{mm}(1) \\ \vdots & \ddots & \vdots & \vdots & \ddots & \vdots \\ 0 & \cdots & 0 & X_{m1}(T) & \cdots & X_{mm}(T) \end{pmatrix} \times \begin{pmatrix} \beta_{11} \\ \vdots \\ \beta_{1m} \\ \vdots \\ \beta_{m1} \\ \vdots \\ \beta_{mm} \end{pmatrix},$$

where $X_{jk}(t) = S_j(t-1)/N_j \cdot \sum_{\tau=1}^{d} P_\tau i_k(t-\tau)$. Denote the matrix with the components $X_{jk}(t)$ by $X$. Then by *structural identifiability* we mean that the following relation holds true:

$$\beta = (X^T X)^{-1} X^T i,$$

where $i = (i_1(1), \ldots, i_1(T), \ldots, i_m(1), \ldots, i_m(T))^T$, and $\beta = (\beta_{11}, \ldots, \beta_{1m}, \ldots, \beta_{m1}, \ldots, \beta_{mm})^T$. Let $B = X^T X$. Structural identifiability of the transmission model holds if and only if the matrix $B$ is non-singular. For instance, consider the case of two age-groups, and note that in such a case $B = \begin{pmatrix} B_1 & 0 \\ 0 & B_2 \end{pmatrix}$, where

$$B_1 = \begin{pmatrix} \sum_{t=1}^{T} X_{11}^2(t) & \sum_{t=1}^{T} X_{11}(t) X_{12}(t) \\ \sum_{t=1}^{T} X_{11}(t) X_{12}(t) & \sum_{t=1}^{T} X_{12}^2(t) \end{pmatrix} \text{ and } B_2 = \begin{pmatrix} \sum_{t=1}^{T} X_{21}^2(t) & \sum_{t=1}^{T} X_{21}(t) X_{22}(t) \\ \sum_{t=1}^{T} X_{21}(t) x_{22}(t) & \sum_{t=1}^{T} X_{22}^2(t) \end{pmatrix}.$$

The determinant of the matrix $B$ is given by $\det(B) = \det(B_1) \times \det(B_2)$ and hence, singularity of $B$ will be determined by the behavior of the matrices $B_1, B_2$. In particular,

$$\det(B_1) = \sum_{t=1}^{T} X_{11}^2(t) \sum_{t=1}^{T} X_{12}^2(t) - \left(\sum_{t=1}^{T} X_{11}(t)X_{12}(t)\right)^2,$$

$$\det(B_2) = \sum_{t=1}^{T} X_{21}^2(t) \sum_{t=1}^{T} X_{22}^2(t) - \left(\sum_{t=1}^{T} X_{21}(t)X_{22}(t)\right)^2.$$

The above equations suggest that if $i_1(t) = i_2(t)$ for all $t$ then the matrix $B$ is singular. In fact, setting $i_1(t) = c \times i_2(t)$ for some constant $c$ for all $t$ leads to singularity of the matrix $B$. Degenerate cases such as $S_1(t) = i_1(t) = 0$ for all $t$ (or other similar combinations, see equations above) also result in the matrix $B$ being singular. In scenarios which are in a close vicinity to a singularity scenario (when $i_1 \approx i_2$), the model, while mathematically identifiable, becomes less identifiable in practice (see SM section I). Due to the complex nonlinearities of the model, it is not straightforward to point out nontrivial scenarios of model parameters for which $i_1 \approx i_2$. Throughout this work we assume to deal with structural identifiable models.

## B. The observation model

We use a measurement error model, in which the observations are given by the deterministic age-group incidence data (equation (1a)) plus an additive normal noise:

(4)  $$Y_j(t) = i_j(t) + \epsilon_j(t) \text{ where } \epsilon_j(t) \sim N(0, \sigma_j(t)^2),$$

where $\sigma_j(t) = \phi_a + \phi_b \cdot i_j(t)$. Here, $\phi_a$ serves as a constant baseline noise and $\phi_b$ is a factor determining the relation between the amount of noise and the incidence on day $t = 1, \ldots, T$.

## III. Methods

## A. A two-stage method for estimating the next-generation matrix

Given observed incidence $Y_j(t)$ for $1 \leq j \leq m$ and $t = 1, \ldots, T$, we wish to estimate the parameters $\beta$ of the NGM, where initially it is assumed that the initial fractions of susceptible in the age-groups ($S_{0,j}$) are known. The estimation can be done using the maximum-likelihood (or nonlinear least squares) approach where one would search for the parameter values that maximize the likelihood function given the observed data (11,26,36). However, the accuracy and computational complexity of maximum-likelihood estimation crucially depends on the initial guess in the parameter space used for optimization, especially when many parameters are considered. Indeed, starting the optimization from random points in the parameter space may lead to local convergence or even lack of convergence of the optimization algorithm (table S3). Furthermore, nonlinear optimization is computationally demanding since one has to "solve" the model for many candidate parameter values (see, e.g. (39) in the context of differential equations). Most of the problems just described can be bypassed by first estimating $\beta$ directly without the use of nonlinear optimization. The direct estimation of model parameters is referred to in the literature as the "direct integral method" in the context of ordinary differential equations (37,40,41). To be more specific, (40) show that the direct integral method is $\sqrt{n}$-consistent ($n$ stands for the sample size) for estimating parameters of fully observed systems of ordinary differential equations linear in functions of the parameters, while (42) prove consistency for partially observed systems. Below, we modify the estimation method to deal with a discrete time model. Since integrals are not needed, we refer to the method as the "direct" method. While the direct method is faster in terms of computational time and circumvents the need for nonlinear optimization, the resulting parameter estimators are not statistically efficient. Thus, this method can be used for generating initial/prior estimates in the parameter space to be used by likelihood estimation, as done here, and hence the two-stage approach. Let $Y$ stand for the vector of observations $\left(Y_1(1), \dots, Y_1(T), \dots, Y_m(1), \dots, Y_m(T)\right)^T$ where each $Y_j(t), t = 1, \dots, T$ is given by equation (4). Then the statistical model takes the matrix form:

(5) $$Y = X\beta + \epsilon$$

where the matrix $X$ and the vector $\beta$ are given in equation (3) and $\epsilon = \left(\epsilon_1(1), \dots, \epsilon_1(T), \dots, \epsilon_m(1), \dots, \epsilon_m(T)\right)^T$ are the measurement errors ($\dim(Y) = mT \times 1, \dim(X) = mT \times m^2, \dim(\beta) = m^2 \times 1, \dim(\epsilon) = mT \times 1$). The least squares solution for $\beta$, i.e., the solution that minimizes the Euclidean norm $\|Y - X\beta\|^2$ is given by:

$$\tilde{\beta} = (X^T X)^{-1} X^T Y.$$

However, in reality $i_j$ is not known and therefore cannot be used when computing $X_{jk}$. Consequently, the matrix $X$ cannot be obtained in order to attain the estimator $\tilde{\beta}$. Hence, we estimate $i_j$ by smoothing the observations $Y_j$. We denote the resulting estimator by $\hat{\imath}_j$. Consequently, we have an estimate of the matrix $X$ denoted by $\hat{X}$ and the resulting estimator for $\beta$ takes the form:

$$\hat{\beta} = \left(\hat{X}^T \hat{X}\right)^{-1} \hat{X}^T \hat{\imath}_j.$$

This is the first step of the estimation procedure, denoted as the 'direct' step. The second step is then to try and further improve the estimate of $\beta$ using maximum-likelihood optimization with the direct estimator $\hat{\beta}$ set as the initial estimate. Based on the observation model (4), the log-likelihood function to maximize is given by:

$$(6) \quad LL(\beta) = -\sum_{j=1}^{m}\sum_{t=1}^{T}\left[\log\left(\sqrt{2\pi}\sigma_j(t)\right) + \frac{\left(Y_j(t)-i_j(t|\beta)\right)^2}{2\sigma_j(t)^2}\right].$$

The final estimator obtained through the second stage of the procedure is thus denoted by $\hat{\beta}_{ML}$. The incidence $i_j(t|\beta)$ is calculated by plugging $\beta$ into the transmission model (1). The likelihood function can be extended to include the parameters $\phi_a$ and $\phi_b$ in case they are unknown. Using the likelihood function above one can also estimate additional unknown parameters such as $\phi_a$, $\phi_b$ and $S_{0_j}$ as explained in the next section.

## B. Estimating the next-generation matrix together with additional model parameters

In many cases, the values of the initial fraction of susceptible in the population $(S_{0_j})$ are unknown. In addition, the transmission model may include more parameters that we would like to estimate such as the effect of weather conditions on the transmission rates (43,44). In such cases, one would need to estimate the additional parameters together with $\beta$. However, note that the transmission model (equation (1)) is linear in parameter $\beta$ but nonlinear in $S_{0_j}$. Thus, while the parameter $\beta$ can be estimated in a 'direct' linear least squares fashion as described above, this is not true with estimating the additional model parameters such as $S_{0_j}$ where nonlinear optimization is unavoidable. We call such a model 'semi-linear in the parameters'. Modification of the direct integral method to models of ordinary differential equations semi-linear in the parameters was studied in (41). Next we suggest a modification for the discrete case of equation (1). Let $\psi$ stand for the set of additional parameter values not including $\beta$ that one would wish to estimate (i.e. $\psi = S_{0_j}$), and consider the matrix $X$ of equation (3). The components $X_{jk}(t)$ of the matrix $X$ will depend on the unknown parameters $\psi$. Thus, in case of a model semi-linear in the parameters we emphasize this dependence using the notation:

$$\beta_\psi = \left(X_\psi^T X_\psi\right)^{-1} X_\psi^T i,$$

which implies that any direct solution $\beta$ depends nonlinearly on additional parameters $\psi$. We estimate $\psi$ via nonlinear maximization of equation (6), where the incidence $i_j(t|\beta_\psi)$ are calculated by plugging $\psi$ and $\beta_\psi$ into the transmission model (1). Once we obtain an estimate $\hat{\psi}$, it is straightforward to obtain an initial estimate for $\beta$ given by:

$$\hat{\beta}_{\hat{\psi}} = \left(\hat{X}_{\hat{\psi}}^T \hat{X}_{\hat{\psi}}\right)^{-1} \hat{X}_{\hat{\psi}}^T \hat{\iota}_j.$$

The second step in the two-stage method in this case is to estimate $\{\psi, \beta\}$ using non-linear maximization of equation (6), while setting the initial parameter estimates to $\{\hat{\psi}, \hat{\beta}_{\hat{\psi}}\}$.

### C. Estimating the next-generation matrix using incidence of recurrent outbreaks

Accurately estimating the NGM $\beta$ together with $S_{0j}$ using incidence data of a single epidemic can be a difficult challenge in some cases (see Results). For an infectious disease with repeated annual epidemics such as influenza, we can improve our estimates by fitting the model to recurrent outbreaks while assuming some model parameters remain fixed over the years. In the case of influenza, a reasonable first assumption regarding the matrix $\beta$ is that it can be broken down to elements that are likely to vary very little over a period of several years and can be approximated as fixed, and elements that do change from year to year but are age-independent. Specifically, age-group contact patterns and the (possibly age-dependent) probabilities of transmission or infection upon contact, should remain more or less the same over a period of several years (this is an unmentioned assumption of most modeling studies employing a contact matrix obtained from surveys as the survey was conducted at one time while the matrix is used to model epidemics at a different time(s) (e.g. (30–36))). However, the transmissibility of the prevailing influenza virus in a specific year could change from year to year (45) but should affect individuals of all age-groups in a similar manner. Thus, we can assume that the NGMs for any two years in this period are the same up to a scaling factor. Denoting $\beta$ as the NGM for year $y = 1$, the NGMs for years $y \geq 2$ are given by $r^y \beta$ for some scaling factors $r^y$. We can then rewrite the transmission model (equation (1)) for recurrent annual outbreaks as follows:

(7a) $$i_j^y(t) = S_j^y(t-1)/N_j^y \cdot \sum_{k=1}^{m}\left(r^y \beta_{jk} \sum_{\tau=1}^{d} P_\tau i_k^y(t-\tau)\right)$$

(7b) $$S_j^y(t) = S_j^y(t-1) - i_j^y(t),$$

where we set $r^1 = 1$. The matrix notation of the observation model for multiple outbreaks can be written using a concatenation of the matrix $X$ and the vector $Y$ in equation (5) for each outbreak, so that for $L$ outbreaks of length $T^y$ ($1 \leq y \leq L$), we obtain $\dim(X) = m \sum_{y=1}^{L} T^y \times m^2$ and $\dim(Y) = m \sum_{y=1}^{L} T^y \times 1$. Here, the components of the matrix $X$ are given by $X_{jk}^y(t) = S_j^y(t-1)/N_j^y \cdot r^y \sum_{\tau=1}^{d} P_\tau i_k^y(t-\tau)$. In this manner, estimating the NGMs requires estimating just $m^2 + L - 1$ parameters instead of $m^2 L$ parameters.

D. Simulation study

Our main goal is to evaluate how well can the NGM $\beta$ be estimated using the two-stage method in each of the scenarios described in sections A-C above. We evaluate the finite sample properties of the two-stage method by simulating incidence data for two age-groups corresponding to children and adults, where the size of the adults age-group was set as two times the size of the children age-group ($N_1 = 10^6$ and $N_2 = 2 \cdot 10^6$). Specifically, we use three 2x2 matrices that can be seen to represent different matrix structures (table 1). In all three matrices the main diagonal elements are larger than the off-diagonal elements as it is expected that more infections are caused by individuals from the same age-group compared to individuals from another age-group. The matrices differ in the relation between the $R_0$ of children and adults (in matrix1 $R_{0_1} > R_{0_2}$, in matrix2 $R_{0_1} = R_{0_2}$ and in matrix3 $R_{0_1} < R_{0_2}$). In addition, matrix2 and matrix3 are symmetrical, implying that a child infects the same number of adults as an adult infects children, while in matrix1 a child infects more adults than an adult infects children. The values of each matrix were set so that $R_0 = 3$ (the spectral radius of the matrix is 3). When estimating $\beta$ using a single outbreak, the initial fraction of susceptible individuals in each age-group were set to $S_{0_1} = S_{0_2} = 0.4$, leading to an effective reproductive number of $R_e = 1.2$ (equation (2)), which is a typical value for seasonal influenza outbreaks (44,45). The initial number of infected in each age-group was set as 0.1% of the initial susceptible population in each age-group, which is a low incidence compared to the unfolding epidemic but high enough so as to avoid any effects of dynamical stochasticity associated with the initial phase of an outbreak (26,46). The serial interval distribution $P_\tau$, $(1 \leq \tau \leq d)$ was set according to the estimated profile for influenza epidemics with $d = 7$ days (26,47).

For each of the examined matrices, age-group incidence data was generated using the transmission model (equation (1)). Given the incidence data, 500 Monte Carlo simulations of the observation model (equation (4)) where ran (verified to be enough simulations in order to obtain similar results upon repeated tests), with the noise parameters set to $\phi_a = 10$ and $\phi_b = 0.1$, producing realistic-looking variance in the observed incidence data (see fig. 1). Using the two-stage method we estimate $\beta$ for each of the simulations, while either assuming $S_{0_j}$ are known (section A), or assuming they are unknown, in which case we estimate them as well using the two-stage method for the semi-linear model (section B). In both scenarios we also estimate the observation model parameters $\phi_a$ and $\phi_b$. We approximate $E(\hat{\beta}_{ML})$ and $Var(\hat{\beta}_{ML})$ using their Monte Carlo counterparts obtained from the 500 estimates of $\beta$ in order to assess the mean square error (MSE) for $\hat{\beta}_{ML}$ given by:

$$(8) \quad MSE(\hat{\beta}_{ML}) = \sum_{j=1}^{m}\sum_{k=1}^{m}\left[\left(E(\hat{\beta}_{ML,jk}) - \beta_{jk}\right)^2 + Var(\hat{\beta}_{ML,jk})\right]$$

In calculating the direct estimator $\hat{\beta}$ we used a 7-day moving average smoothing to obtain $\hat{i}_j$. The non-linear maximization of equation (6) was performed using the simplex algorithm (48) as it was implemented by the fminsearch function in Matlab (R2016a).

To test the effect of fitting multiple epidemics data using the assumption described in section C, we generated ten years of incidence data using the same three matrices. We selected values for $R_e$ and $S_{0\,j}$ for each year $y = 1,\ldots,10$ such that $avg(R_e^y) = 1.2$, $std(R_e^y) \approx 0.05$, $avg\left(S_{0\,j}^y\right) = 0.4$ and $std\left(S_{0\,j}^y\right) \approx 0.05$ (see table S1 for details). These values give a year to year variation in the simulated outbreak sizes that resembles the year to year variation observed in actual influenza incidence data (compare simulated data to real outbreaks data in fig. S2 and fig. 4, respectively). Given the values for $R_e^y$, $S_{0\,j}^y$ and the NGM $\beta$, it is possible to calculate the scaling factor $r^y$ that provides the NGM for year $y$, $\beta^y = r^y \beta$:

$$(9) \quad R_e^y = \rho\begin{pmatrix} \beta_{11}^y S_{0\,1}^y & \beta_{12}^y S_{0\,1}^y \\ \beta_{21}^y S_{0\,2}^y & \beta_{22}^y S_{0\,2}^y \end{pmatrix} = \rho\begin{pmatrix} \frac{\beta_{11}}{r^y} S_{0\,1}^y & \frac{\beta_{12}}{r^y} S_{0\,1}^y \\ \frac{\beta_{21}}{r^y} S_{0\,2}^y & \frac{\beta_{22}}{r^y} S_{0\,2}^y \end{pmatrix} \rightarrow r^y = R_e^y / \rho\begin{pmatrix} \beta_{11} S_{0\,1}^y & \beta_{12} S_{0\,1}^y \\ \beta_{21} S_{0\,2}^y & \beta_{22} S_{0\,2}^y \end{pmatrix}$$

The factors $r^y$, calculated for each matrix $\beta$ (table S1), were used with the transmission model (equation (7)) to generate incidence data for multiple outbreaks. As before, we ran a Monte Carlo study, generating 500 stochastic simulations of ten outbreaks using the observation model (equation (4)). We fit these simulated data for two scenarios: i) assuming $S_{0\,j}^y$ ($y = 1,\ldots,10$) are known and ii) assuming they are unknown. In both cases, we estimate the parameters of $\beta$ together with the additional parameters $r^y$ for $y = 2,\ldots,10$ (in order to force identifiability in this scenario we fixed $r^1$ to its actual value and did not estimate it), by employing the two-stage method for the semi-linear model. In these scenarios, we fix the observation model parameters $\phi_a$ and $\phi_b$ to their actual values and did not attempt to estimate them, as estimating these parameters together with the numerous transmission model parameters proved to be a computationally demanding and time consuming task to perform as part of a Monte Carlo simulation study.

### E. Estimating the next-generation matrix from influenza-like-illness incidence data

We employ the two-stage method in order to fit influenza-like-illness (ILI) incidence data in two age-groups of children (0-19) and adults (20+). The ILI data depicts diagnoses given to patients visiting the community clinics of the Maccabi health maintenance organization (HMO) in Israel, which has a nationwide coverage of 20%-25% (table S6). The incidence data includes eleven seasonal influenza epidemics between 1998 and 2013. We excluded three seasons from this dataset – the 2002/2003 and 2005/2006 seasons, which were dominated by an influenza B virus as opposed to an influenza A virus in the rest of the seasons, and the 2009/2010 season, which was dominated across the globe by a novel influenza A/H1N1 strain (generating an influenza pandemic). The unique circumstances of these three seasons present themselves in the incidence data with features that are not observed in the rest of the seasons (45). In order to properly fit these three seasons together with the rest of the seasons we need to enhance the model to incorporate more details (e.g. the effect of weather conditions and school vacations on the transmission rates) (36,44), which was out of the scope for the current paper. We determined the beginning and ending of each influenza season according to the findings of virological tests for influenza virus performed on suspected ILI patients visiting sentinel clinics of the Israeli Health Ministry (see table S6). The daily ILI incidence was smoothed using a 7-day moving average, in order to deal with the 'weekend effect' (the incidence dropping close to zero during weekends as clinics are closed). We used the following observation model to fit the ILI data:

(10) $$ILI_j^y(t) = \eta_j^y \theta_j^y i_j^y(t) + \epsilon_j^y(t),$$

where $ILI_j^y$ is the ILI incidence in age-group $j$ at year $y$, $i_j^y$ is the influenza incidence in age-group $j$ at year $y$ (given by the transmission model, equation (7)), $\eta_j^y$ are the surveillance rates or the fraction of Maccabi members in the whole population and $\theta_j^y$ are the reporting rates which represents the relation between actual influenza incidence and the ILI incidence in the whole population (see table S6 for information regarding the surveillance and reporting rates). $\epsilon_j^y$ are the observations errors, which were modeled as was done in observation model (4), $\epsilon_j^y(t) \sim N\left(0, \sigma_j^y(t)^2\right)$ where $\sigma_j^y(t) = \phi_a + \phi_b \cdot \eta_j^y \theta_j^y i_j^y(t)$. By fitting the ILI data, the NGM for each seasonal epidemic outbreak is estimated, assuming that the matrix remains the same between years up to a scaling factor. This is accomplished by estimating a single NGM $\beta$ for all years and a scaling factor $r^y$ for $y = 2, \ldots, 11$ with $r^1$ (i.e. the first year) fixed to 1, so that the NGM for year $y$ is $\beta^y = r^y \beta$ (see section C above for details). We also estimate the initial fraction of susceptible individuals in each of the age-groups for each year $S_{0j}^y$, as well as the observation model parameters $\phi_a$ and $\phi_b$. The initial number of infected in each age-group is set according to the ILI incidence in the week prior to the start of the fit. We employ the two-stage method for semi-linear model to obtain estimates for the model parameters. We calculate 95% confidence intervals (CI) for the estimated NGM components using profile likelihood (11) after testing the method on the simulated data and verifying its 95% nominal coverage (see fig. S4).

## IV. Results

### A. Estimating the next-generation matrix from simulated data

We first present the results of exploring the finite sample properties of the two-stage method used for estimating the NGM $\beta$ based on simulated data. As detailed in the Methods (section III(D)), three 2x2 matrices were employed under four different scenarios:

i. fitting incidence of a single epidemic assuming the initial fraction of susceptible population in each age-group ($S_{0_j}$) are known

ii. fitting incidence of a single epidemic while estimating $S_{0_j}$

iii. fitting incidence of ten recurrent epidemics while estimating the scaling factors for the NGM in year $y$ ($r^y, y > 1$), assuming $S_{0_j}^y$ are known

iv. fitting incidence of ten recurrent epidemics while estimating $r^y$ and $S_{0_j}^y$

Figure 2 summarizes the estimates for $\beta$ obtained by fitting the simulated data using only the first step of the method (the direct estimates, see also table S2) and using the complete two-stage method (direct + maximum-likelihood, see also table 1). In addition, we examined the results of fitting the simulated data using maximum-likelihood starting from random initial values (table S3). Based on the simulation study we note the following findings:

1. When fitting a single outbreak, the MSE (equation (8)) is gradually growing from matrix1 to matrix2 and then to matrix3. A possible explanation for this finding is that the ability to estimate $\beta$ is related to the similarity (in the sense discussed above) of the incidence functions of the two groups. This may result in the matrix $B = X^T X$ being closer and closer to singular, leading to higher variance of the parameter estimators (see SM section I).

2. The estimates of $\beta$ from a single outbreak are much more accurate when $S_{0_j}$ are known (scenario (i)) compared to when $S_{0_j}$ are unknown (scenario(ii)). In all three matrices, the bias and variance obtained from the Monte Carlo simulations are considerably smaller in scenario (i) compared to scenario (ii) (see scenarios (i) and (ii) in fig. 2 and table 1).

3. Using the assumption regarding $\beta$ being constant up to a factor from year to year allows to obtain much more accurate estimates for $\beta$ when fitting ten outbreaks with varying initial conditions. When $S^y_{0_j}$ are known (scenario (iii)), $\beta$ can be estimated almost perfectly (the bias and variance obtained from the Monte Carlo simulations are close to zero - see scenario (iii) in fig. 2 and table 1). Even when $S^y_{0_j}$ are unknown (scenario (iv)), $\beta$ can be estimated rather accurately. The same is also true for estimates of $S^y_{0_j}$ (fig. 3 and table S4) and $r^y$ (table S5). The estimates improve when fitting ten outbreaks together compared to fitting a single outbreak. Note that with matrix1 and matrix2 the variance in the estimates of $S_0$ for adults is larger than in the children. We conjecture that due to the nonlinearity of the model the variance of the estimators is a function of the model parameters, however, verifying this theoretically or numerically is beyond the scope of this work. It is expected that with an increasing number of outbreaks fitted, the obtained MSE will decrease further.

4. Using maximum-likelihood optimization starting from the estimates given by the direct method improves the estimates of $\beta$ (compare table 1 and table S2). In one case we did obtain a lower MSE using the direct estimates compared to the maximum-likelihood estimates (matrix3, scenario (ii)). In this case, the direct estimates had smaller variance than the maximum-likelihood estimates at the expanse of a larger bias, as can be expected due to smoothing (40). The results obtained using maximum-likelihood optimization while starting from random initial values are almost always worse than the results obtained by starting the maximum-likelihood optimization from the direct method estimates (see table S3).

## B. Estimating the next-generation matrix from ILI data

Figure 4 shows the best fit obtained to the ILI data using the two-stage method estimates. As explained in the Methods (section III(E)), the NGM for each year is obtained by estimating a matrix for the first year and a multiplicative factor for each of the following years that captures the relation between the matrix in year $y > 1$ and the first year matrix. The estimated NGM for the first year (the 1998 season) obtained by this fit is $\begin{pmatrix} 2.05\ [1.97-2.13] & 0.11\ [0.08-0.14] \\ 0.30\ [0.26-0.33] & 1.85\ [1.76-1.94] \end{pmatrix}$, with the 95% CI for each component of the matrix obtained using profile likelihood given in parenthesis (see fig. S3). The estimated multiplicative factors for the NGMs in the following years are $r^{2,\dots,11} = 1.39, 1.14, 1.53, 1.27, 1.96, 2.33, 1.65, 1.88, 0.88$ and $1.25$. The obtained matrix is highly assortative, indicating that most of the transmission occurs within the two age-groups, with the highest transmission being from children to children and the lowest transmission from adults to children. According to this matrix, in an entirely susceptible population a child would infect, on average, 1.2 times more individuals than an adult would, as $\frac{R_{0\,children}}{R_{0\,adults}} = \frac{2.05+0.30}{0.11+1.85} \approx 1.2$. Figure 5 shows the estimated $S_0$, $R_0$ and $R_e$ values for each age-group and the population as a whole, in each of the years (see also table S6). For most seasons, the estimated $S_0$ values for children are higher than those of the adults. In 2010, a year after the arrival of the novel pandemic strain, $S_0$ in both age-groups is estimated to be much higher than previous years, with a larger fraction of susceptible children compared to adults. A year later, in 2011, the estimates of $S_0$ in both age-groups are back to their 'normal' value range. The estimates for the effective reproductive number ($R_e$) in the population as a whole vary from 1.15 to 1.50 with considerable differences in $R_e$ between children and adults. The values of $R_e$ in adults are typically below or just above the threshold of one, indicating that without children, seasonal influenza epidemics would not manage to spread in the population (at least not every year). The maximum-likelihood estimates for the variance components are $\hat{\phi}_a = 12.6$ and $\hat{\phi}_b = 0.05$.

## V. Discussion

The above analysis demonstrates the potential of using incidence data to obtain good estimates of both the NGM, as well as the population susceptibility, without the need to make prior assumptions on the structure of the matrix. By implementing a large simulation experiment, several aspects of the finite sample properties of the two-stage method used for estimating the NGM were recognized. It was demonstrated that given incidence data of a single outbreak, the NGM can be accurately estimated, if the initial fractions of susceptible population are known (e.g., when it is known that the whole population is susceptible). When the initial population susceptibility is unknown, estimating the NGM becomes more difficult, but still feasible. We have shown that it is possible to improve the estimates using incidence from recurrent outbreaks. This is performed by assuming that age-group transmission patterns remain fixed over a period of several years, while the transmissibility of the pathogen ($R_0$) may vary between years but affect individuals of all age-groups in a similar manner. Our analysis and simulations reveal that as the age-groups incidence curves are similar to each other (in the sense that $i_2 \approx c \times i_1$), it becomes more difficult to establish parameter identifiability. This suggests, that when deciding if and how to partition incidence data of the whole population into age-group incidence data, considering the distinction in the incidence of the age-groups in the suggested partitioning should play a major role. However, this empirical observation requires further theoretical analysis of the model in order to understand it better.

Estimating the parameters of the age-group model was not a trivial task. Indeed, although maximum-likelihood estimation has desirable statistical properties, at present, there is no clear scheme (either numerical or analytical) that can guarantee optimal parameter estimates. The likelihood function in our case is multivariable, and may possibly have complex surfaces with numerous local minima, maxima and saddle points that may lead to local convergence of optimization algorithms. In particular, finding the maximum-likelihood estimate can be vastly reliant on the selected initial conditions used in the optimization procedure, thus making the pursuit for adequate optimum computationally demanding and complex. Therefore, our analysis employed a two-stage approach for the estimation procedure, which in step one uses a direct method to obtain initial parameter estimates that are then plugged into a maximum-likelihood optimization in step two. The effect of the direct step can be seen when examining the results of maximum-likelihood optimization starting from random initial parameter values, which lead to estimates with much higher MSE compared to the results of the two-stage method (table S3). One can attempt to improve the results by starting the optimization from numerous sets of random initial values. This might be feasible when estimating a small number of parameters. However, with a growing number of parameters to estimate, it becomes increasingly difficult to converge to the optimum values using this method, and to do so in a reasonable amount of time. The direct step helps the optimization avoid local solutions and reach the optimum values faster, and appears to be more robust in some of the scenarios we explored. In addition, the linear (in the parameters) representation of the model enables to obtain some basic identifiability properties of the model and hence, a better understanding of the problem at hand.

The statistical properties of the direct estimator will depend on the amount of smoothing of the observations, which leads to the typical bias-variance trade off (see e.g., (40) in the context of ordinary differential equations). Deriving the theoretical properties of the direct estimator is outside the scope of this work. In the simulation study performed here, the observations were smoothed using a 7-day moving average. However, when fitting the ILI data, we did examine the effect of the size of the smoothing window by running the two-stage procedure using different smoothing windows for the first step and comparing the results obtained at the end of the second step. We found that while the size of the smoothing window had some effect on the estimates obtained at the first step, the effect did not carry over to the estimates obtained at the end of the second step. That is, the estimates of the two-stage method were not sensitive to the size of the smoothing window used for the direct estimator (a similar observation was made in (37)).

Previous attempts of estimating the NGM using incidence data were restricted to the initial phase of the outbreak. Given such data limitations forced either making simplifying assumptions about the structure of the matrix so as to reduce the number of estimated parameters (25), or combine incidence data with contact tracing data to bypassed the need to restrict the matrix structure (26). The current work demonstrates that given enough high quality incidence data it is possible to obtain reliable estimates of the NGM without the need to restrict the matrix or use contact data. We found that the Israeli NGM for influenza is highly assortative, implying that most of the infections are restricted to within the age-groups, with the maximum transmission rates focused within the children. This finding is consistent with the POLYMOD study, which estimated that the strongest levels of contacts were between children as well (29). Similar matrix structure was also obtained in a previous attempt to estimate the NGM for pandemic influenza (26). The parameter estimates of the key relevant epidemiological parameters $S_0$, $R_0$ and $R_e$ for the population as a whole, obtained by fitting the age-group model to eleven years of ILI data from Israel, are in line with previous values estimated using a simple SIR model with no age-groups (44,45). More so, the qualitative findings of (45) who observed an increasing trend in $R_0$ and a decreasing trend in $S_0$ between 1998-2009, were also obtained in this study using the more complex age-group model (fig. 5). However, by fitting an age-group model we also obtain estimates for $S_0$, $R_0$ and $R_e$ in each age-group, which can be of importance. For example, the fact that the estimates of $R_e$ in adults are close to or below unity implies that targeted interventions towards children might lead to the prevention of seasonal influenza outbreaks altogether. Similar results were obtained for pandemic influenza in Israel (36) and for seasonal influenza in other locations across the globe (49–52). In general, estimates of epidemiological parameters of an age-group model can be helpful in implementing and improving vaccine allocation and other mitigation strategies used to reduce the burden of infectious diseases.

One possible difficulty regarding the use of incidence data in fitting age-group models is that in many cases records are incomplete, due for instance, to partial detection (44,53). However, application of the methods presented here does not necessitate the use of complete incidence data. As long as the reporting rate of the different age-groups is constant in time and/or is not age dependent, the data can be used (see (25,47) for further discussion). Differences in the reporting rate among different age-groups (for example, if members of some age-groups are less likely to seek medical care) will lead to biases in the estimated NGM. Therefore, it is crucial to have a good assessment of the reporting rates in the different age-groups, possibly through some independent sources. Here, we established the relation between ILI and influenza in children and adults based on the results of (36), using the estimated influenza incidence obtained by fitting virological and serological data collected during the 2009/10 pandemic (see table S6), but more information is necessary to verify how good are these assessments for seasonal influenza epidemics. Consideration needs to be given as well to the possibility of changes in the reporting rates during the outbreak (34,36), resulting in a change in the fraction of cases detected, which can also lead to biases in the obtained estimates.

In a future research, we intend to further test the capability of the two-stage method to obtain accurate estimates of the NGM in more complex models (e.g., additional age-groups). There is also a need for further analysis to improve our understanding on the issue of structural identifiability. Finally, one would like to derive theoretical properties of the direct approach, used in the first step of the two-stage method, such as consistency, convergence rates, and limiting distribution which in turn, will lead to better understanding of the problem and the estimation methods.

## Acknowledgements

The authors would like to thank Professor Guy Katriel, Professor Brian Everitt and two referees for their insightful comments which helped to improve the paper and the presentation of the results.

## Funding

This research was supported by the Israeli Science Foundation grant no. 387/15, and by a Grant from the GIF, the German-Israeli Foundation for Scientific Research and Development number I-2390-304.6/2015.

## Declaration of Conflicting Interests

The authors declare that there is no conflict of interest.

## References

1. Anderson RM, May RM. Infectious Diseases of Humans: Dynamics and Control. Oxford University Press; 1992.

2. Keeling MJ, Rohani P. Modeling Infectious Diseases in Humans and Animals. Princeton university press. Princeton university press; 2007. 408 p.

3. Grassly NC, Fraser C. Mathematical models of infectious disease transmission. Nat Rev Micro. 2008;6(june):477–87.

4. Huppert A, Katriel G. Mathematical modelling and prediction in infectious disease epidemiology.

Clin Microbiol Infect. 2013 Jul 30;19(June):999–1005.

5. Heesterbeek H, Anderson RM, Andreasen V, Bansal S, De Angelis D, Dye C, et al. Modeling infectious disease dynamics in the complex landscape of global health. Science (80- ). 2015;347(6227):aaa4339-aaa4339.

6. Ferguson NM, Donnelly CA, Anderson RM. The foot-and-mouth epidemic in Great Britain: pattern of spread and impact of interventions. Science (80- ). American Association for the Advancement of Science; 2001;292(5519):1155–60.

7. Keeling MJ, Woolhouse MEJ, May RM, Davies G, Grenfell BT. Modelling vaccination strategies against foot-and-mouth disease. Nature. 2003;421(6919):136–42.

8. Lipsitch M, Cohen T, Cooper B, Robins JM, Ma S, James L, et al. Transmission dynamics and control of severe acute respiratory syndrome. Science. 2003 Jun;300(5627):1966–70.

9. Earn D, He D, Loeb M, Fonseca K. Effects of School Closure on Incidence of Pandemic Influenza in Alberta, Canada. Ann Intern Med. 2012;173–82.

10. Yaari R, Kaliner E, Grotto I, Katriel G, Moran-Gilad J, Sofer D, et al. Modeling the spread of polio in an IPV-vaccinated population; lessons learned from the 2013 silent outbreak in Southern Israel. BMC Med. 2016;14(1):95.

11. Bolker BM. Ecological Models and Data in R. New Jersy: Princeton university press; 2008.

12. Morton A, Finkenstadt BF. Discrete time modelling of disease incidence time series by using Markov chain Monte Carlo methods. J R Stat Soc Ser C (Applied Stat. 2005 Jun;54(3):575–94.

13. Ionides EL, Bretó C, King a a. Inference for nonlinear dynamical systems. Proc Natl Acad Sci U S A. 2006 Dec;103(49):18438–43.

14. Hooker G, Ellner SP, Roditi LDV, Earn DJD. Parameterizing state-space models for infectious disease dynamics by generalized profiling: measles in Ontario. J R Soc Interface. 2011;8(60) :961–74.

15. Kretzschmar M, Teunis PFM, Pebody RG. Incidence and reproduction numbers of pertussis: Estimates from Serological and Social Contact Data in Five European Countries. PLoS Med. 2010;7(6).

16. Baguelin M, Hoschler K, Stanford E, Waight P, Hardelid P, Andrews N, et al. Age-Specific Incidence of A/H1N1 2009 Influenza Infection in England from Sequential Antibody Prevalence Data Using Likelihood-Based Estimation. Viboud C, editor. PLoS One. 2011 Feb;6(2):e17074.

17. Ott JJ, Stevens GA, Groeger J, Wiersma ST. Global epidemiology of hepatitis B virus infection: New estimates of age-specific HBsAg seroprevalence and endemicity. Vaccine. 2012;30(12):2212–9.

18. Anderson, R. M. & May RM. The control of communicable diseases by age-specific immunization schedules. Lancet. 1982;319(160).

19. Schenzle D. An Age-Structured Model of Pre- and Post-Vaccination Measles Transmission. IMA J Math Appl Med Biol. 1984;1:169–91.

20. Diekmann O, Heesterbeek JAP, Metz JAJ. On the definition and the computation of the basic

reproduction ratio R0 in models for infectious diseases in heterogeneous populations. J Math Biol. 1990;28(4):365–82.

21. Diekmann, Heesterbeek J. Mathematical Epidemiology of Infectious Diseases: Model Building, Analysis and Interpretation - O. Diekmann, J. A. P. Heesterbeek. Wiley Series. 2000. 322 p.

22. Kanaan MN, Farrington CP. Matrix models for childhood infections: a Bayesian approach with applications to rubella and mumps. Epidemiol Infect. 2005 Jun;133(6):1009.

23. Van Effelterre T, Shkedy Z, Aerts M, Molenberghs G, Van Damme P, Beutels P. Contact patterns and their implied basic reproductive numbers: an illustration for varicella-zoster virus. Epidemiol Infect. Cambridge Univ Press; 2009;137(1):48–57.

24. Klepac P, Pomeroy LW, Bjørnstad ON, Kuiken T, Osterhaus ADME, Rijks JM. Stage-structured transmission of phocine distemper virus in the Dutch 2002 outbreak. Proc R Soc London B Biol Sci. The Royal Society; 2009;rspb--2009.

25. Glass K, Mercer GN, Nishiura H, McBryde ES, Becker NG. Estimating reproduction numbers for adults and children from case data. J R Soc Interface. 2011 Sep 7;8(62):1248–59.

26. Katriel G, Yaari R, Huppert A, Roll U, Stone L. Modelling the initial phase of an epidemic using incidence and infection network data: 2009 H1N1 pandemic in Israel as a case study. J R Soc Interface. 2011 Jan 19;8(59):856–67.

27. Edmunds WJ, O'Callaghan CJ, Nokes DJ. Who mixes with whom? A method to determine the contact patterns of adults that may lead to the spread of airborne infections. Proc Biol Sci. 1997 Jul;264(1384):949–57.

28. Wallinga J, Teunis P, Kretzschmar M. Using data on social contacts to estimate age-specific transmission parameters for respiratory-spread infectious agents. Am J Epidemiol. 2006 Nov;164(10):936–44.

29. Mossong J, Hens N, Jit M, Beutels P, Auranen K, Mikolajczyk R, et al. Social contacts and mixing patterns relevant to the spread of infectious diseases. PLoS Med. 2008 Mar;5(3):e74.

30. Ogunjimi B, Hens N, Goeyvaerts N, Aerts M, Van Damme P, Beutels P. Using empirical social contact data to model person to person infectious disease transmission: An illustration for varicella. Math Biosci. 2009;218(2):80–7.

31. Rohani P, Zhong X, King AA. Contact network structure explains the changing epidemiology of pertussis. Science. 2010;330(6006):982–5.

32. Merler S, Ajelli M, Pugliese A, Ferguson NM. Determinants of the spatiotemporal dynamics of the 2009 h1n1 pandemic in europe: Implications for real-time modelling. PLoS Comput Biol. 2011;7(9).

33. Fumanelli L, Ajelli M, Manfredi P, Vespignani A, Merler S. Inferring the Structure of Social Contacts from Demographic Data in the Analysis of Infectious Diseases Spread. PLoS Comput Biol. 2012;8(9):35–9.

34. Eames KTD, Tilston NL, Brooks-Pollock E, Edmunds WJ. Measured dynamic social contact patterns explain the spread of H1N1v influenza. PLoS Comput Biol. 2012 Jan;8(3):e1002425.

35. Goeyvaerts N, Willem L, Van Kerckhove K, Vandendijck Y, Hanquet G, Beutels P, et al. Estimating

dynamic transmission model parameters for seasonal influenza by fitting to age and season-specific influenza-like illness incidence. Epidemics. 2015;13:1–9.

36. Yaari R, Katriel G, Stone L, Mendelson E, Mandelboim M, Huppert A. Model-based reconstruction of an epidemic using multiple datasets: understanding influenza A/H1N1 pandemic dynamics in Israel. J R Soc Interface. 2016;13(116):92–92.

37. Dattner I. A model-based initial guess for estimating parameters in systems of ordinary differential equations. Biometrics. Wiley Online Library; 2015;71(4):1176–84.

38. Katriel G. Stochastic discrete-time age-of-infection epidemic models. Int J Biomath. 2013;6(4).

39. Voit EO, Almeida J. Decoupling dynamical systems for pathway identification from metabolic profiles. Bioinformatics. 2004;20(11):1670–81.

40. Dattner I, Klaassen CAJ, others. Optimal rate of direct estimators in systems of ordinary differential equations linear in functions of the parameters. Electron J Stat. The Institute of Mathematical Statistics and the Bernoulli Society; 2015;9(2):1939–73.

41. Dattner I, Miller E, Petrenko M, Kadouri DE, Jurkevitch E, Huppert A. Modelling and parameter inference of predator– prey dynamics in heterogeneous environments using the direct integral approach. J R Soc Interface. 2017;

42. Vujačić I, Dattner I. Consistency of direct integral estimator for partially observed systems of ordinary differential equations linear in the parameters. arXiv Prepr arXiv160205761. 2016;

43. Shaman J, Pitzer VE, Viboud C, Grenfell BT, Lipsitch M. Absolute humidity and the seasonal onset of influenza in the continental United States. PLoS Biol. 2010;8(2).

44. Yaari R, Katriel G, Huppert A, Axelsen JB, Stone L. Modelling seasonal influenza: the role of weather and punctuated antigenic drift. J R Soc Interface. 2013;10(84):1–12.

45. Huppert A, Barnea O, Katriel G, Yaari R, Roll U, Stone L. Modeling and Statistical Analysis of the Spatio-Temporal Patterns of Seasonal Influenza in Israel. PLoS One. 2012;7(10).

46. King AA, Domenech de Celles M, Magpantay FMG, Rohani P. Avoidable errors in the modelling of outbreaks of emerging pathogens, with special reference to Ebola. Proc R Soc B Biol Sci. 2015;282:20150347–20150347.

47. Roll U, Yaari R, Katriel G, Barnea O, Stone L, Mendelson E, et al. Onset of a pandemic: characterizing the initial phase of the swine flu (H1N1) epidemic in Israel. BMC Infect Dis. 2011;11(1):92.

48. Lagarias JC, Poonen B, Wright MH. Convergence of the restricted Nelder-Mead algorithm in two dimensions. Optimization. 2011;48109:27.

49. Longini IM, Halloran ME. Stragegy for distribution of influenza vaccine to high-risk groups and children. Am J Epidemiol. 2005;161(4):303–6.

50. Bansal S, Pourbohloul B, Meyers LA. A comparative analysis of influenza vaccination programs. PLoS Med. 2006;3(10):1816–25.

51. Dushoff J, Plotkin JB, Viboud C, Simonsen L, Miller M, Loeb M, et al. Vaccinating to protect a vulnerable subpopulation. PLoS Med. 2007;4(5):0921–7.

52. Baguelin M, Flasche S, Camacho A, Demiris N, Miller E, Edmunds WJ. Assessing optimal target populations for influenza vaccination programmes: an evidence synthesis and modelling study. PLoS Med. 2013 Oct;10(10):e1001527.

53. Barnea O, Huppert A, Katriel G, Stone L. Spatio-temporal synchrony of influenza in cities across Israel: The "Israel is one city" hypothesis. PLoS One. 2014;9(3):1–11.

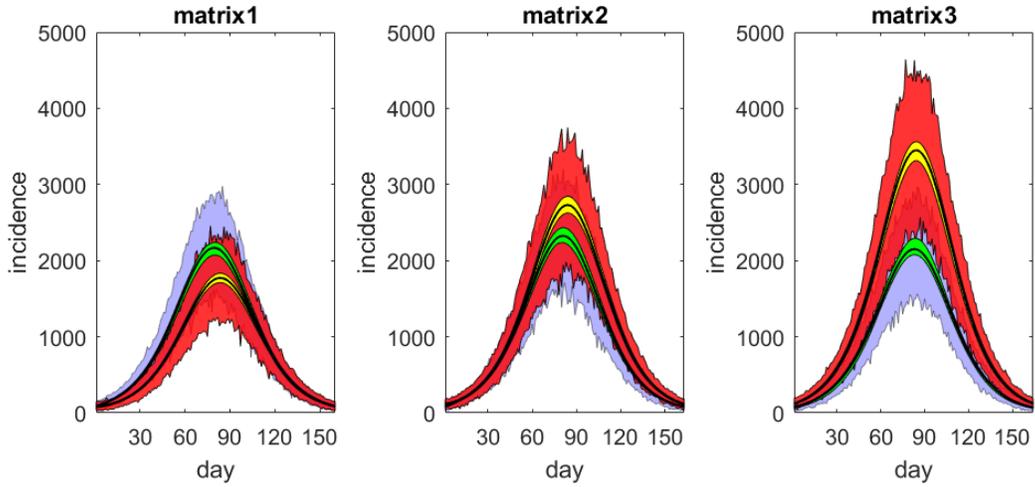

**Figure 1**: Simulated age-group incidence for two age-groups (children and adults) obtained using three different NGMs termed matrix1, matrix2 and matrix3 (see table 1 for parameter values). The adult population was set as twice the size of the children population. Black curves are showing the incidence obtained using the deterministic transmission model (equation (1)), with the matrices set so that $R_0 = 3$ and the initial fraction of susceptible individuals in both age-groups set to $S_0 = 0.4$. Blue (children) and red (adults) regions show the incidence obtained from 500 simulations of the stochastic observation model (equation (4)) with $\phi_a = 10$ and $\phi_b = 0.1$. Green (children) and yellow (adults) regions show the fit obtained to the stochastic simulations using the two-stage method while estimating $S_{0_j}$ as well (scenario (ii)).

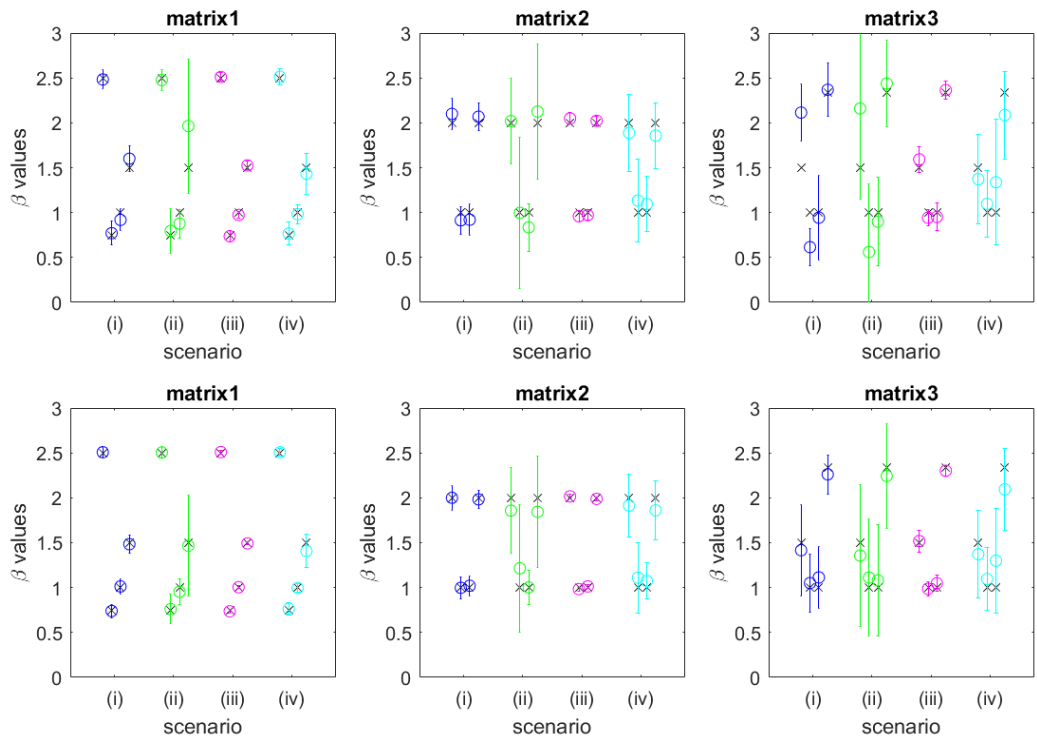

**Figure 2**: Summary of fitting three 2x2 NGMs to simulated incidence data using only the direct method (top panels, see also table S2) and using the full two-stage method (bottom panels, see also table 1). 'x' marks the actual values of the NGMs. 'o' and bars mark the mean and standard deviation obtained by fitting 500 instances of the stochastic observation model (equation (4)). The four scenarios are:

(i) fitting the incidence of a single epidemic assuming $S_{0_j}$ are known (blue)

(ii) fitting the incidence of a single epidemic while estimating $S_{0_j}$ as well (green)

(iii) fitting the incidence of ten epidemics assuming $S_{0_j}^y$ are known (magenta)

(iv) fitting the incidence of ten epidemics while estimating $S_{0_j}^y$ as well (cyan)

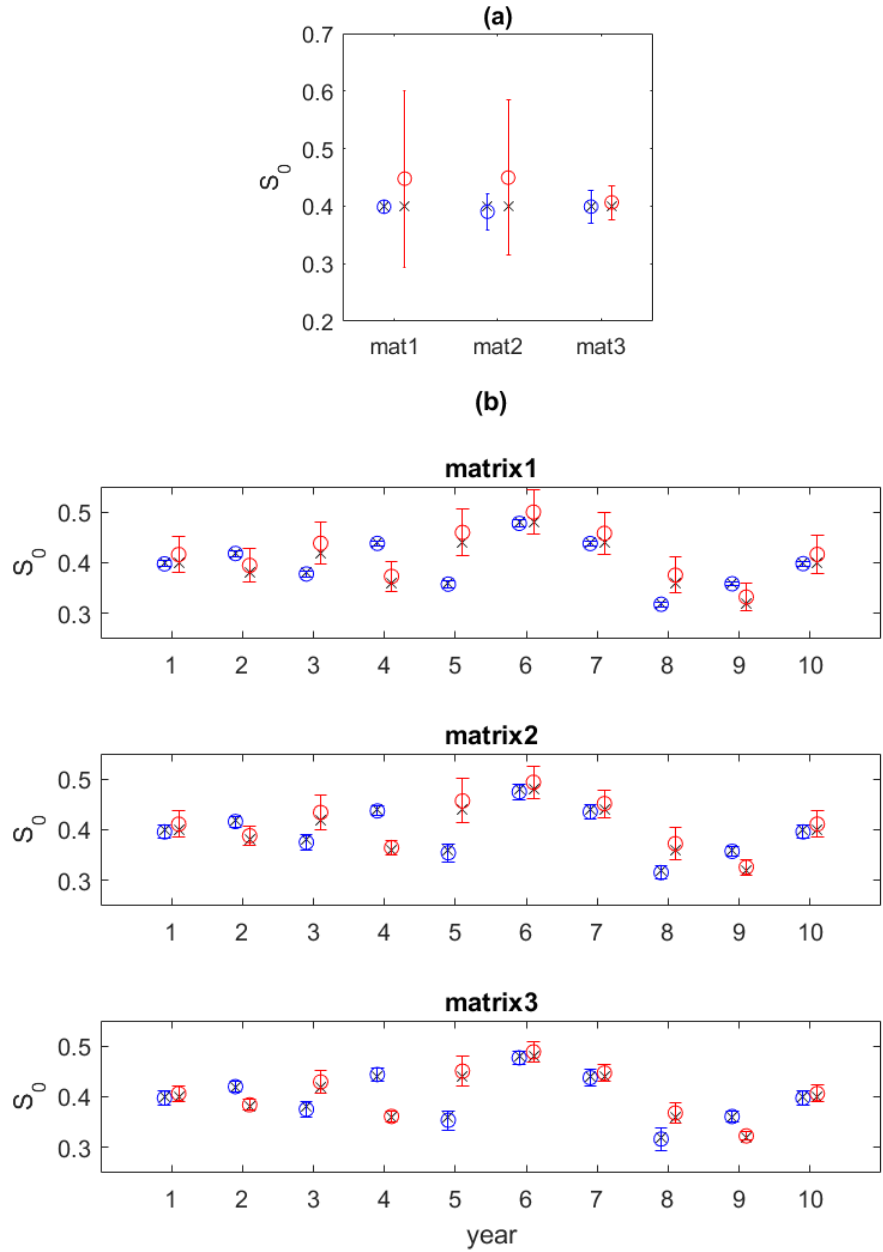

**Figure 3**: Mean estimates and standard deviation of $S_0$ for children (blue) and adults (red) obtained by fitting 500 instances of the stochastic observation model (equation (4)) using the two-stage method. 'x' marks the actual values, while 'o' represents the mean estimates (see also table S4).

(a) Estimates for each NGM, obtained by fitting a single outbreak (scenario (ii))

(b) Estimates per year for each NGM, obtained by fitting ten recurrent outbreaks (scenario (iv))

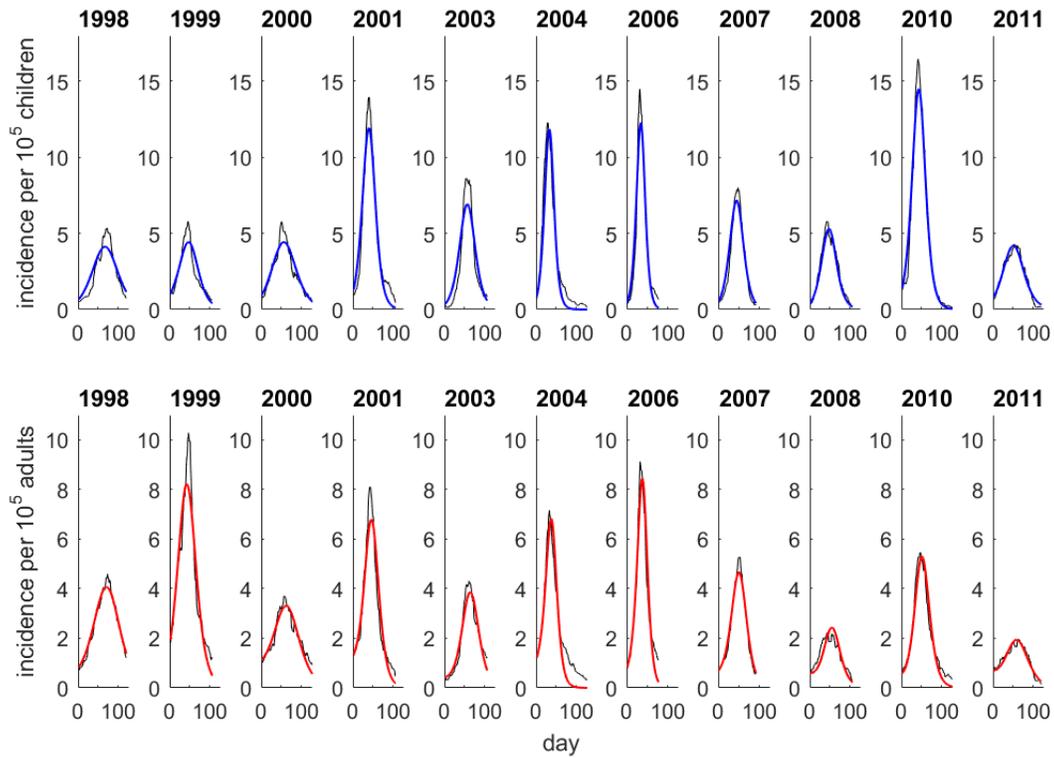

**Figure 4**: Fit to eleven years of ILI data (after 7-days moving average smoothing) from Israel in two age-groups: children 0-19 (top panels) and adults 20+ (bottom panels), obtained by fitting observation model (10) using the two-stage method. Black curves show the ILI data while the blue/red curves show the obtained fit.

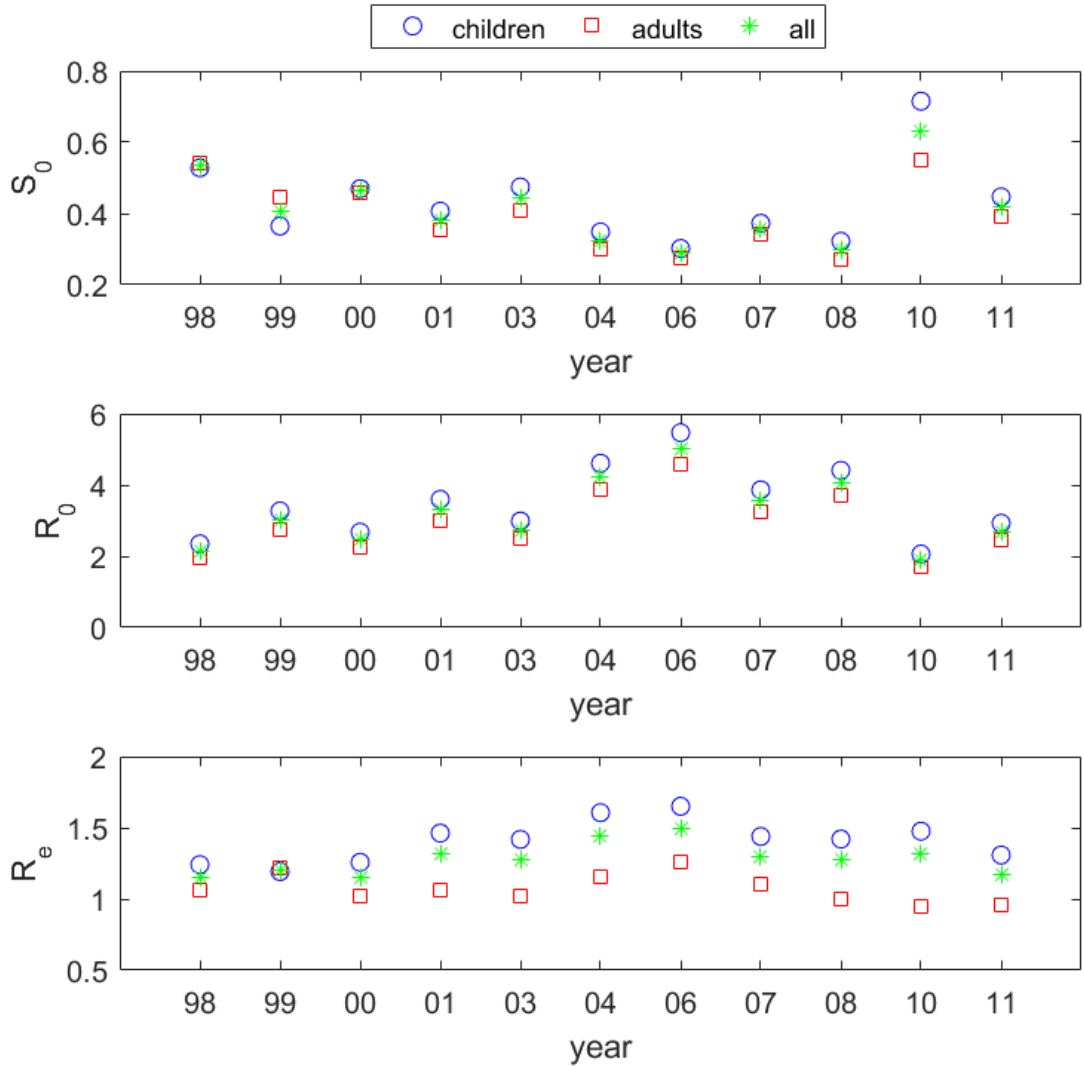

**Figure 5**: Estimates of $S_0$, $R_0$, and $R_e$ per year in children, adults and the population as a whole (see also table S6). The basic reproductive number ($R_0$) in year $y$ and age-group $j$ is calculated as $R_{0j}^y = r^y \cdot \sum_{l=1}^{m} \beta_{lj}$, and in the population as a whole $R_0^y = \rho(r^y \cdot \beta)$. Similarly, the effective reproductive number ($R_e$) is calculated for each age-group and the population as a whole using the fraction of susceptible individuals ($S_0$) in each age-group (equation (2)).

|  | scenario (i) (MSE = 0.027) | | | scenario (ii) (MSE = 0.354) | | | scenario (iii) (MSE = 0.002) | | | scenario (iv) (MSE = 0.053) | | |
|---|---|---|---|---|---|---|---|---|---|---|---|---|
| matrix1 | mean | bias | std. | mean | bias | std. | mean | bias | std. | mean | bias | std. |
| $\beta_{11}$=2.50 | 2.51 | 0.01 | 0.06 | 2.51 | 0.01 | 0.06 | 2.51 | 0.01 | 0.02 | 2.51 | 0.01 | 0.04 |
| $\beta_{12}$=0.75 | 0.74 | -0.01 | 0.07 | 0.76 | 0.01 | 0.16 | 0.74 | -0.01 | 0.02 | 0.76 | 0.01 | 0.07 |
| $\beta_{21}$=1.00 | 1.01 | 0.01 | 0.09 | 0.96 | -0.04 | 0.14 | 1.00 | 0 | 0.02 | 1.00 | 0 | 0.05 |
| $\beta_{22}$=1.50 | 1.48 | -0.02 | 0.10 | 1.47 | -0.03 | 0.55 | 1.49 | -0.01 | 0.02 | 1.41 | -0.09 | 0.19 |
| matrix2 | scenario (i) (MSE = 0.057) | | | scenario (ii) (MSE = 1.467) | | | scenario (iii) (MSE = 0.003) | | | scenario (iv) (MSE = 0.468) | | |
|  | mean | bias | std. | mean | bias | std. | mean | bias | std. | mean | bias | std. |
| $\beta_{11}$=2.00 | 2.00 | 0 | 0.14 | 1.81 | -0.19 | 0.53 | 2.02 | 0.02 | 0.02 | 1.91 | -0.09 | 0.35 |
| $\beta_{12}$=1.00 | 1.00 | 0 | 0.12 | 1.28 | 0.28 | 0.76 | 0.98 | -0.02 | 0.02 | 1.11 | 0.11 | 0.39 |
| $\beta_{21}$=1.00 | 1.02 | 0.02 | 0.11 | 1.02 | 0.02 | 0.21 | 1.01 | 0.01 | 0.02 | 1.07 | 0.07 | 0.20 |
| $\beta_{22}$=2.00 | 1.98 | -0.02 | 0.10 | 1.88 | -0.12 | 0.64 | 1.99 | -0.01 | 0.02 | 1.86 | -0.14 | 0.33 |
| matrix3 | scenario (i) (MSE = 0.559) | | | scenario (ii) (MSE = 1.801) | | | scenario (iii) (MSE = 0.036) | | | scenario (iv) (MSE = 1.080) | | |
|  | mean | bias | std. | mean | bias | std. | mean | bias | std. | mean | bias | std. |
| $\beta_{11}$=1.50 | 1.42 | -0.08 | 0.51 | 1.37 | -0.13 | 0.78 | 1.52 | 0.02 | 0.12 | 1.37 | -0.13 | 0.49 |
| $\beta_{12}$=1.00 | 1.05 | 0.05 | 0.32 | 1.10 | 0.10 | 0.65 | 0.99 | -0.01 | 0.08 | 1.10 | 0.10 | 0.35 |
| $\beta_{21}$=1.00 | 1.11 | 0.11 | 0.35 | 1.11 | 0.11 | 0.62 | 1.05 | 0.05 | 0.09 | 1.30 | 0.30 | 0.58 |
| $\beta_{22}$=2.33 | 2.26 | -0.07 | 0.22 | 2.23 | -0.10 | 0.58 | 2.30 | -0.03 | 0.06 | 2.10 | -0.23 | 0.46 |

**Table 1**: Results of estimating the parameters of three NGMs by fitting the transmission model (equations (1) or (7)) to 500 Monte-Carlo simulations of the observation model (equation (4)), using the two-stage method. The results are given for the following four scenarios:

(i) fitting the incidence of a single epidemic assuming $S_{0_j}$ are known

(ii) fitting the incidence of a single epidemic while estimating $S_{0_j}$

(iii) fitting the incidence of ten recurrent epidemics assuming $S_{0_j}^y$ are known

(iv) fitting the incidence of ten recurrent epidemics while estimating $S_{0_j}^y$